\documentclass[journal]{IEEEtran}
\ifCLASSINFOpdf
  \usepackage[pdftex]{graphicx}
\else
\fi
\usepackage{caption}
\usepackage{amsmath}
\usepackage{algorithm, algpseudocode}

\begin{document}
%
\title{Providing reliability and auditability to the IoT LwM2M protocol through Blockchain}
%
%
%

\author{Cristian~Mart\'in, Iv\'an Alba, Joaqu\'in Trillo, Enrique Soler, Bartolom\'e~Rubio and Manuel~D\'iaz
\thanks{Cristian~Mart\'in, Enrique Soler, Bartolom\'e~Rubio and Manuel~D\'iaz are with the ITIS software institute at the University of M\'alaga, Spain. Iv\'an Alba and Joaqu\'in Trillo are with the Department of Languages and Computer Science at the University of M\'alaga, M\'alaga, Spain; e-mail: Cristian Mart\'in (cmf@lcc.uma.es).}}

%
%

\markboth{Cristian Mart\'in \MakeLowercase{\textit{et al.}}: Providing reliability and auditability to the LwM2M protocol through Blockchain, ArXiv, August~2020}%
{Shell \MakeLowercase{\textit{et al.}}: Bare Demo of IEEEtran.cls for IEEE Journals}
%



\maketitle

\begin{abstract}
Blockchain has come to provide transparency, reliability as well as to increase the security in computer systems, especially in distributed ones like the Internet of Things (IoT). A few integrations have been proposed in this context so far; however, most of these solutions do not pay special attention to the interoperability of the IoT, one of the biggest challenges in this field. In this paper, a Blockchain solution has been integrated into the OMA Lightweight M2M (LwM2M), a promising industry IoT protocol for global interoperability. This integration provides reliability and auditability to the LwM2M protocol enabling IoT devices (LwM2M clients) to transparently interact with the protocol. Furthermore, a missing reliable API to allow users and applications to securely interact with the system and an interface to store critical information like anomalies for auditability have been defined.
\end{abstract}


%
\IEEEpeerreviewmaketitle

\section{Introduction}
\label{sec:introduction}
It is unquestionable that the Internet of Things (IoT) \cite{gubbi2013internet} is offering an unprecedented revolution to society. The IoT ability to reduce the gap between the physical world and the digital one allows the automation and optimization of multiple processes and services in addition to providing a better knowledge of physical phenomena. The possibilities offered by this field are uncountable as demonstrated daily by industrial, domestic and institutional examples such as Industry 4.0, connected health or smart cities.

Centralized architectures like the ones used in cloud computing have significantly contributed to the development of IoT applications \cite{diaz2016state}. However, regarding data transparency they act as black boxes and network participants do not have a clear vision of how and where the information they provide is going to be used. In this regard, Blockchain reduces the reliability and security problems of centralized environments, while improving its adaptation to merely distributed systems such as the IoT \cite{reyna2018blockchain}. Blockchain is based on the concept of ``chain of blocks'', in which information is grouped into sets (blocks), which are interconnected with each other through a timeline in such a way that, thanks to cryptographic techniques and consensus protocols, it is not possible to edit or repudiate this information without modifying the entire timeline, something that is costly in computational terms. 

The development of IoT applications involves the management of IoT devices, which may include their identification, registration, credential management and access control  to applications. Despite their high availability, centralized architectures like cloud environments can suffer transparency and reliability issues and, in addition to this, present a single point of failure due to their centralized nature. Furthermore, transparency and reliability can also be required to the IoT produced information itself, especially when a secure proof of facts is needed. As seen in other mission-critical systems like black boxes in airplanes, the recording of events is of critical importance to reconstruct their whole sequence, especially when a disaster occurs. In the IoT field, this is not well extended yet; however, there may be situations where it could be adequately incorporated. For instance, in Structural Health Monitoring \cite{alonso2018middleware} it might be worth having the log of events that led into a bridge or a tunnel collapse to study its origin and to be able to identify the root cause. On the other hand, open standards are the only way to ensure IoT interoperability, which is overwhelmed with many protocols and solutions. In this regard, most of the solutions tend to provide vertical silos which limit even more the adoption of the IoT in the society.

To overcome these challenges, in this work we present an integration between the IoT and Blockchain, a system that provides secure and reliable management of IoT devices, applications and critical information. The system architecture has been designed to have a minimal impact on the IoT devices, which are limited per se. Through an accessible Web interface, users can easily manage the credentials of IoT devices and applications. An Application Programming Interface (API) has been defined to provide authorized applications the ability to register immutable and tamper-proof critical events. 

This work has been built over the Lightweight Machine-to-Machine (LwM2M) \cite{LwM2M} standard. LwM2M was defined by the Open Mobile Alliance (OMA) to satisfy the need for a device management protocol, semantics for resource identification and access, and end-to-end security in the Industry 4.0. LwM2M has been built on top of popular IoT Constrained Application Protocol (CoAP) and the IETF stack and is intended to be the de facto solution for global interoperability in the IoT. This protocol enables the management of IoT devices, firmware updates and applications to acquire sensing information through a standard specification. As discussed in \cite{bertin2019access}, LwM2M provides a straightforward solution for access control, but can be limited to handle dynamic IoT environments. This protocol has been extended in this work to provide reliable management of device and application credentials and critical information through Blockchain, which also enables the secure mobility of devices between LwM2M deployments.

Therefore, the main contributions of this paper are:

\begin{enumerate}
    \item Extend the LwM2M protocol to enable reliable and auditable management of IoT device credentials in Blockchain.
    
    \item Incorporate a secure and reliable API in LwM2M to manage the access control of IoT applications through Blockchain.
    
    \item Provide a reliable service to register critical IoT information for auditability.
    
\end{enumerate}

The rest of the paper is organized as follows. Section \ref{sec:background} presents a background on Blockchain and LwM2M. Related work is discussed in Section \ref{sec:related-work}. In Section \ref{sec:architecture} the system architecture and its components are presented. The system implementation is detailed in Section \ref{sec:implementation} and evaluation is presented in Section \ref{sec:evaluation}.  Lastly, our conclusions and future work are presented in Section \ref{sec:conclusions-future}.

\section{Background}
\label{sec:background}
\subsection{LwM2M}
\label{subsec:lwm2m}

Lightweight M2M (LwM2M) is an Open Mobile Alliance (OMA) protocol that provides a light and fast solution for managing M2M and IoT devices. LwM2M aims to meet the growing market demand for M2M solutions in the Industry 4.0. It is specially designed to achieve a reduction in energy and data consumption, ideal for low-capacity devices and networks that require efficient use of bandwidth.

The LwM2M protocol architecture operates under the client-server paradigm extending the Constrained Application Protocol (CoAP) \cite{coap}. Unlike CoAP, LwM2M clients offer resources in IoT devices that are securely accessible and managed in a standardized way through LwM2M servers. LwM2M defines four interfaces for communication between clients and servers:
\begin{enumerate}
\item Bootstrap
\item Client Registration
\item Device Management and Service Enablement
\item Information Reporting
\end{enumerate}

The protocol defines a simple model where client information is organized as resources that can be accessed through the previous interfaces. Resources are organized into different objects within the client, and each object can have multiple instances. Each object and resource has a unique identifier. For example, objects 0 and 1 are assigned to LwM2M security and the connection with LwM2M server respectively, and 
/1/0/8 is the Registration Update Trigger resource in the instance 0 of the resource 1 (connection with LwM2M server). 

The Bootstrap interface is provided by the Bootstrap servers and manages the credentials and bootstrapping to the different LwM2M clients so that they can later register to the LwM2M servers. The Client Registration interface registers LwM2M clients on LwM2M servers. Three operations are provided by this interface: registering an LWM2M client on an LWM2M server, updating the registration (this operation allows to have a control of connected LwM2M clients), and de-registering to delete a client registration on an LwM2M server.

The Device Management and Service Enablement interface allows LwM2M servers to access the object instances and resources available on the LwM2M clients connected to them. This interface performs the interaction between clients and servers through the use of operations such as Read, Write and Execute.

\begin{figure*}[h]
    \centering
    \includegraphics [scale=1]{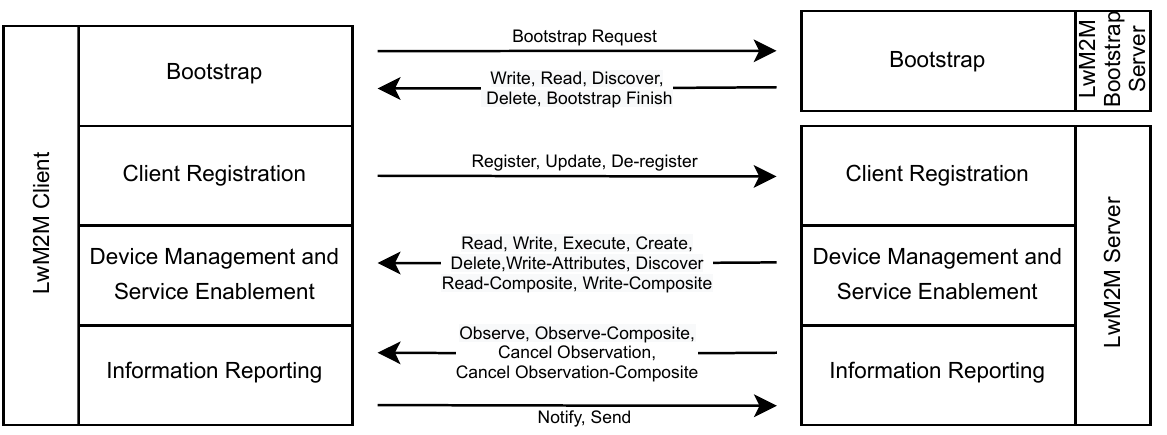}
    \caption{LwM2M interfaces and workflow}
    \label{fig:LwM2M_Interface}
\end{figure*}

Finally, the Information Reporting interface allows LwM2M clients to report asynchronously information about any changes in their resources to LwM2M servers. The observation is initialized by the Observe operation from LwM2M servers to an object, resource or instance of an LwM2M client and it is maintained until the operation is canceled. Fig. \ref{fig:LwM2M_Interface} shows all the interactions between a client and an LwM2M server through the operations of these four interfaces. 

The LwM2M enabler uses CoAP and therefore UDP links, but it can also use a SMS link. Security over UDP is provided by DTLS (Datagram Transport Layer Security).

There are different implementations of LwM2M. Currently, the most used are Wakaama \cite{wakaama} and Leshan \cite{leshan}. Wakaama is a C implementation of LwM2M clients and servers. It is designed to be portable on POSIX compliant systems. To test the capabilities of the Bootstrap server, some samples and tests are available in its official repository. Wakaama belongs to the open source community of the Eclipse Foundation. On the other hand, Leshan is a Java implementation of LwM2M servers and clients. Leshan is based on a CoAP implementation Californium and the DTLS implementation Scandium. Like Wakaama, it belongs to the Eclipse foundation and some samples and tests are also available in its official repository. In addition, a web UI (user interface) for both Bootstrap and LwM2M servers and a public test sandbox with a Bootstrap server and an LwM2M server are available so that its functionality can be tested quickly and easily.

\subsection{Blockchain}
\label{subsec:blockchain}
At any digital system, the trust in the incoming and outcoming information is one of its biggest concerns. And this situation is even more challenging when no verification mechanisms are provided, particularly when the system has to deal with sensitive data. In 2008, two innovative concepts appeared in this context and they were introduced by Satoshi Nakamoto \cite{bitcoin}: Bitcoin and Blockchain. Bitcoin is the well-known first cryptocurrency, a purely peer-to-peer version of electronic cash. This system allows people to make payments without a third party intermediary like a bank or any other financial institution. The second concept presented by Nakamoto, Blockchain, is the mechanism under the hood of Bitcoin. And nowadays its popularity and applicability are larger than the cryptocurrency itself.

Blockchain is defined by Bashir \cite{bashir} as ``a peer-to-peer, distributed ledger that is cryptographically secure, append-only, immutable, and updateable only via consensus or agreement among peers''. A Blockchain is a distributed ledger of a chronological chain of records in the form of encrypted blocks made up by all transactions executed by the participants. Initially, Blockchain was only the distributed ledger system for the Bitcoin cryptocurrency but currently is being researched and applied in many areas as financial services, supply chain and logistics, healthcare, IoT, smart cities, among other applications.

There are three types of Blockchain networks: public, private and consortium. The last two types are also called permissioned. Public Blockchain has absolutely no restrictions and are open to the public; anyone can send transactions, run applications and join the network. This kind of network utilizes some type of a Proof of Work (PoW) or Proof of Stake (PoS) algorithm as consensus protocol. Bitcoin is the most famous public Blockchain. Another well-known public Blockchain network is Ethereum \cite{ethereum}. According to its inventor, Vitalik Buterin, Ethereum is ``a Blockchain with a built-in fully fledged Turing-complete programming language that can be used to create `contracts' ... simply by writing up the logic in a few lines of code''. And this is the main difference with Bitcoin: Ethereum Blockchain allows smart contracts.

A smart contract is a computer program that encapsulates source code and the business logic necessary to execute a function when certain conditions are met. The concept of smart contract was defined by Nick Szabo at the end of the last century \cite{szabo1996}. Before Blockchain, Szabo's idea could not become a reality, and so it was only a definition/concept. With the advent of this technology, the first smart contracts could be created. Not all Blockchains support smart contracts. For example, Bitcoin does not allow to code this kind of computer program.

Permissioned blockchains are Blockchain networks where access is controlled by one or more organizations. As explained previously, private and consortium blockchains are permissioned. The main difference between them is that private blockchains are managed by one single organization while consortium blokchains are controlled by more than one party. Permissioned blockchains are more popular within industry and business areas for which role definition, security and identity are important. The most famous example of permissioned Blockchain is the Hyperledger project \cite{hyperledger}. Hyperledger is not a project by itself; it is an open-source umbrella project composed by frameworks, libraries and tools for enterprise-grade Blockchain developments. Started in December 2015 by the Linux Foundation, Hyperledger collaborates with more than 250 companies (IBM, Inter, SAP Ariba, among others) and has six Blockchain framework projects. The best-known project is Hyperledger Fabric \cite{hyperledgerfabric}, an enterprise-grade permissioned distributed ledger written in Go. At the beginning of 2019 the first long-term-support version of Fabric was released.

\section{Related work}
\label{sec:related-work}
The IoT and Blockchain are mainly decentralized and distributed \cite{khan2018iot}, and this is maybe why they complement very well each other: by providing real world information and enabling decentralized, reliable and auditable authentication and access control to IoT devices respectively.

In \cite{novo2018blockchain} and \cite{novo2018scalable} an architecture for managing decentralized roles and permissions in the IoT through Blockchain is proposed. Although LwM2M is also adopted as backbone IoT protocol, this architecture requires the communication of IoT devices with a new entity known as Management Hub, which is responsible for the interaction with the Blockchain network. In our architecture, IoT devices and external users and applications just interact with LwM2M components (Bootstrap and LwM2M servers) as proposed by the specification, therefore the flow in the LwM2M standard has not been altered. The architecture presented in this paper aims at improving the reliability and auditability on the LwM2M standard in a transparent way to IoT devices and end users and applications. Moreover, centralizing all the communications through the Management Hub can also have a negative impact to resource-constrained IoT devices.

The authors in  \cite{biswas2018scalable} present a Blockchain framework for secure and scalable transactions in the IoT. The fundamental principle of the framework is that IoT devices do not interact directly with a Blockchain peer, but with an intermediate entity in a Local Peer (Lpeer) network, which comprises a local ledger that restricts the number of transactions entering the global Blockchain. An Lpeer network also includes a Certificate Authority (CA) that provides authentication and registration for IoT devices and an Lpeer node that enables the interaction with the main Blockchain. Although this work speeds the transactions in IoT networks, the authentication and authenticity of devices and users are guaranteed by CAs that can be secured but may present the weakness of centralized architectures.

Limitations on IoT devices can suppose a barrier to apply directly Blockchain on them. To overcome that, many works go to edge and fog computing as intermediary entity between the IoT and Blockchain. In \cite{ren2019identity} a Blockchain access control management based on edge computing is presented. A self-certified public key-based system is used for the registration and authentication of IoT devices, and the identities and certificates are stored in a Blockchain network composed of edge centers. Thanks to the Blockchain network, a trust relationship between edge centers and IoT devices can be established. Access control rights can be transferred from one edge center to another, thus enabling IoT mobility between edge networks. In \cite{guo2019blockchain} a hierarchical Blockchain architecture for IoT authentication based on edge computing is also presented.  However, as also happens in BSeIn \cite{lin2018bsein}, these works merely focus on authentication and control access and do not provide an interoperable and standard protocol to interact with IoT devices like the one presented in this paper.

In \cite{almadhoun2018user}, a user authentication scheme using Blockchain-enabled fog nodes to IoT devices is proposed. This work also adopts Ethereum smart contracts network and has the source code available on GitHub. Users authenticate directly in the Blockchain network, and in case of success, a token is generated, which with a public key could authenticate with fog servers to interact with IoT devices. Finally, users can establish a normal secure SSL connection with IoT devices. The main drawback of this solution is the high requirements imposed by SSL connections, specially for resource-constrained devices.

In \cite{hammi2018bubbles}, a decentralized authentication method for IoT devices based on Blockchain is proposed. In this work the concept ``bubbles of trust'' is introduced, in which secure virtual zones are created where only authorized devices can communicate each other in a secure way. The dependency of a public Blockchain and the consequences of this in terms of cryptocurrencies and latency are addressed by \cite{cui2020hybrid} by providing an authentication scheme through a hybrid Blockchain in wireless sensors networks. Although these works are not based on a open standard for IoT interactions, they do provide a fine control of inter-device interactions that will be taken into account as future work.

BlockBDM \cite{zhaofeng2019blockchain} is a hierarchical Blockchain architecture for big data management and secure consumption. The architecture is organized as follows: 1) a permissioned Blockchain enables trust management, security proof of data sources and history usage; 2) a public token-based Blockchain encourages users to provide high-quality content obtaining economic profit; 3) and finally, InterPlanetary File System (IPFS), a Blockchain data storage, is adopted to store big data (e.g., multimedia or remote sensing image) in a reliable, decentralized and flexible way. In general, BlockBDM provides a good solution for data management, sharing and accountability in big data systems. The solution presented in this paper focuses on IoT applications that require device and user management in a standard IoT protocol and, at the same time, enables the recording of critical events for forensics.

Unlike Sensor-Chain \cite{shahid2019sensor}, Tornado \cite{liu2019tornado} and other approaches to enabling Blockchain in IoT, our work just uses Blockchain for data accountability and reliability as a service. Despite the promising future of this integration and the continuous advances made to reduce power consumption and adapt Blockchain to the IoT, this work has adopted a Blockchain-as-a-service approach because: 1) IoT devices do not notice the presence of Blockchain, since they communicate with Bootstrap and LwM2M servers and interact through them with the Blockchain network; 2) the no presence of a Blockchain allows to remove its dependency and facilitates the adaption of the system to already running deployments; and finally 3) in spite of the improvements and relaxed consensus protocols, enabling Blockchain in the IoT has still an overhead which may seriously reduce the life expectancy of battery-powered IoT nodes.

\section{Reliable and auditable architecture for IoT LwM2M networks}
\label{sec:architecture}

\begin{figure*}[h]
    \centering
    \includegraphics [scale=0.7]{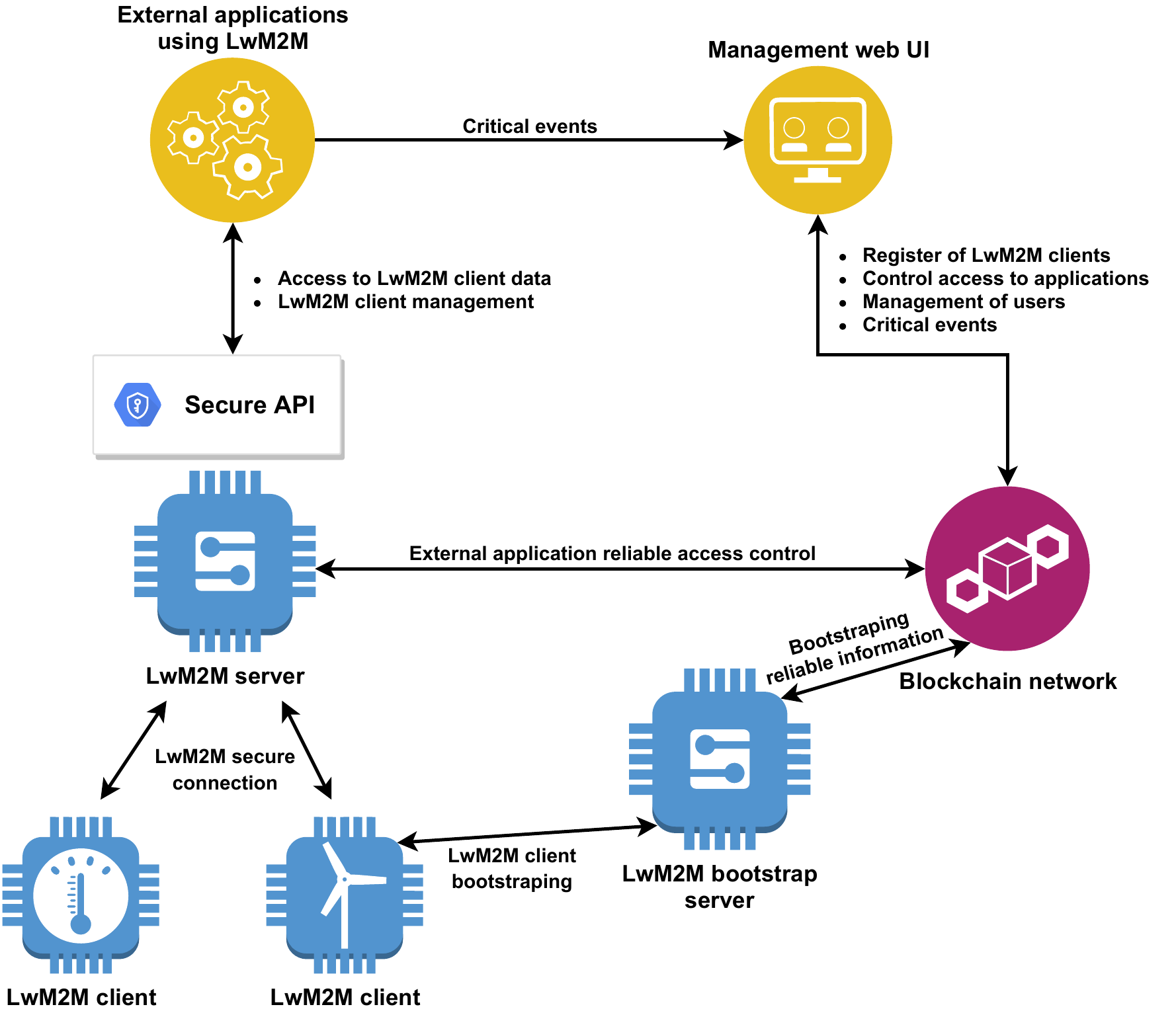}
    \caption{Overall reliable and auditable architecture for IoT LwM2M networks}
    \label{fig:architecture}
\end{figure*}

The system presented in this paper offers not only reliable and auditable management of IoT devices and applications, but also the ability to record critical events when required for later data forensics. This is of special interest to reconstruct the whole sequence of events during a critical occurrence like a natural disaster and for auditability. Through this system, whose architecture is shown in Fig. \ref{fig:architecture}, the management of IoT devices, applications and critical events can be auditable in a reliable way through Blockchain. Next, each component of the architecture is detailed.

\subsection{Blockchain network and smart contracts}
\label{subsec:arch-blockchain}

As previously explained, the recording of critical events needs to be reliable and auditable. And to achieve this, these critical events must be stored in a secure way for later forensics. This is the main reason, along with the authentication management in IoT devices, to use Blockchain in the proposed architecture. As described in subsection \ref{subsec:blockchain}, there are three main types of Blockchain networks: public, consortium and private. For this architecture, we have adopted a public Blockchain since the system needs to be world-wide reliable and not only be trusted by one organization or within a set of partners (private or consortium network, respectively).

The architecture also requires the definition of smart contracts. These smart contracts are open to anyone (because they are deployed in a public Blockchain network), but the interaction with them must be limited. Only certain users could interact with the defined contracts, executing transactions on them. The system needs three different smart contracts: one to register LwM2M clients; another to store critical information for auditability; and the last one to manage users and applications who can use the Management Web UI and the APIs of the different LwM2M servers deployed respectively.

\subsection{Management web UI}
\label{subsec:arch-management-app}
\begin{figure}[h]
    \centering
    \includegraphics [scale=0.65]{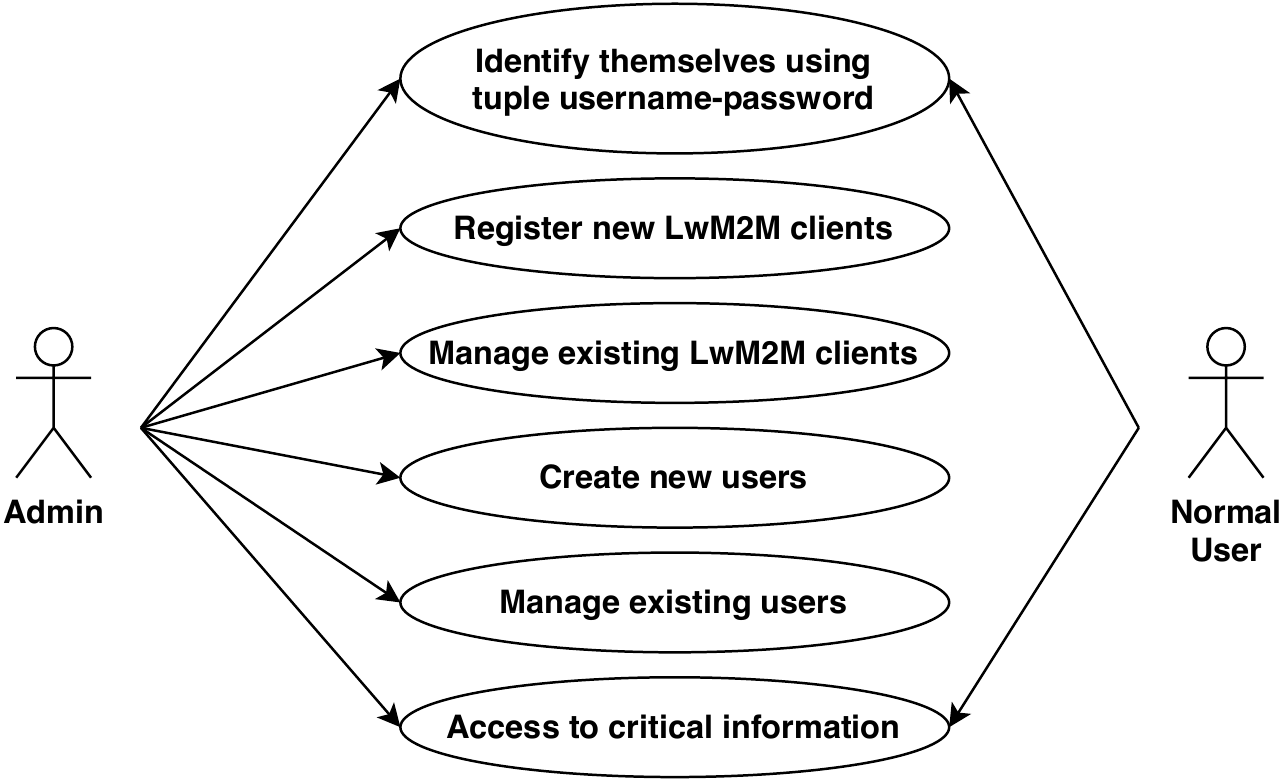}
    \caption{Management Web Application use cases}
    \label{fig:useCases}
\end{figure}

To manage the system, an accessible and user-friendly management web UI has been defined and implemented. There are two different roles in the  management Web: admin and normal user. On the one hand, users with the admin role can create user and device credentials. They can also manage existing users and devices and visualize all existing critical information stored in Blockchain like anomalies. On the other hand, normal users can only access to the critical information stored in smart contracts at the Blockchain network. Fig. \ref{fig:useCases} summarizes all the described use cases depending on the user role.

The management web UI also provides an interface where external users and applications can store critical information like anomalies on the Blockchain network. Note that this can suppose the management web UI to act as a centralized system, however, it can be distributedly deployed in different LwM2M deployments to solve this. With this, only authorized and authenticated applications and users can store and access critical information.

\subsection{LwM2M Bootstrap servers}
\label{subsec:arch-lwm2m-bootstrap-servers}
As described in subsection \ref{subsec:lwm2m}, Bootstrap servers manage the credentials and bootstrapping to the different LwM2M clients so that they can later register into the LwM2M servers.

\begin{figure*}[h]
    \centering
    \includegraphics [scale=0.8]{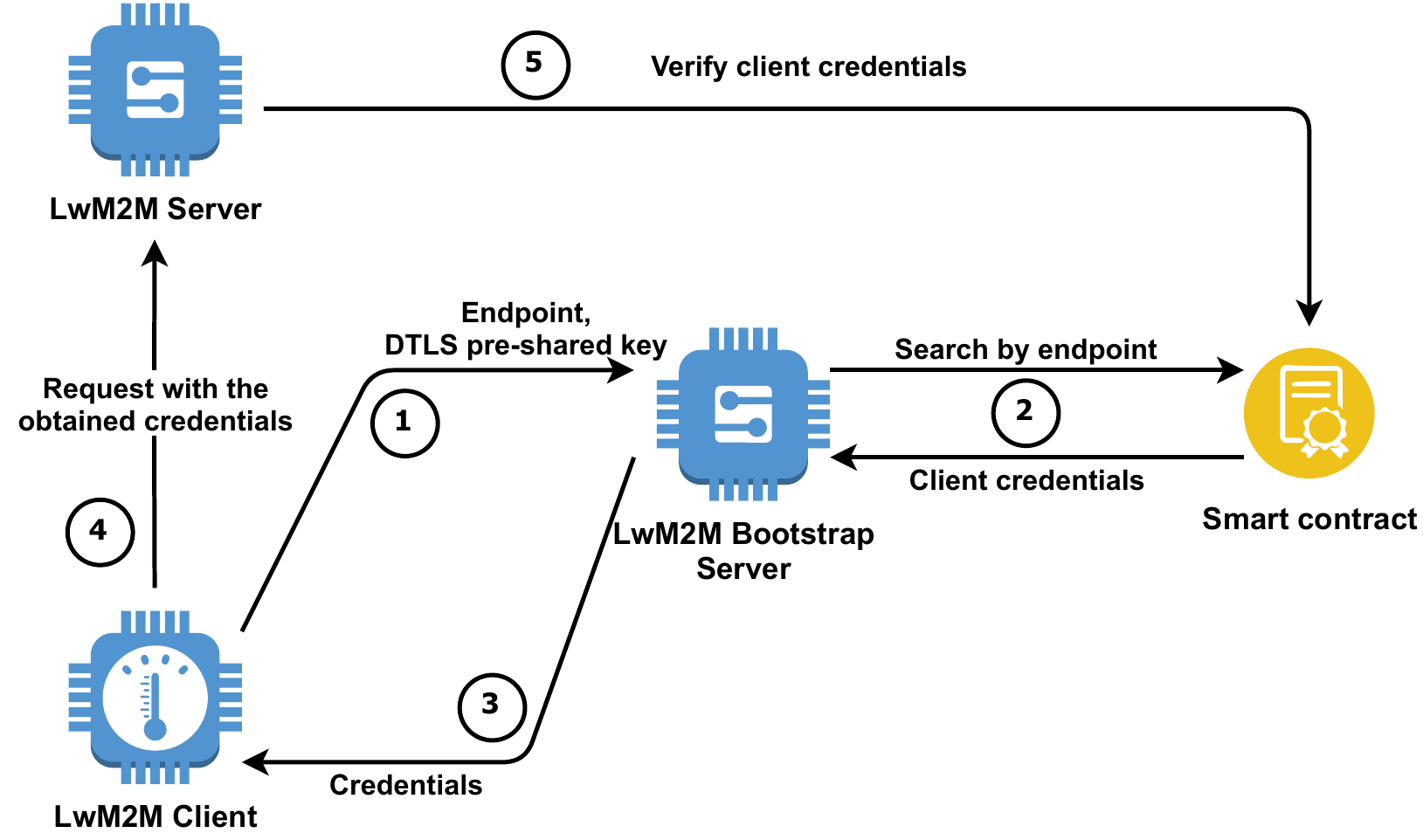}
    \caption{Data flow of the client authentication process on the Bootstrap server}
    \label{fig:deviceIdentification}
\end{figure*}

In this system, we have modified the Bootstrap servers so they can access the Blockchain network to query all the necessary client information for authentication and registration. A smart contract (ClientStore contract) request is made every time a client's information needs to be obtained. This interaction with the Bootstrap server is shown in Fig. \ref{fig:deviceIdentification}.

Clients authenticate to the Bootstrap server using a pre-shared key (PSK), the authentication scheme supported for the time being by this system. The Bootstrap server consults in the ClientStore contract the information of the clients that make the request. This search is performed by the registered endpoint name. Once the client is successfully authenticated, the Bootstrap server provides the necessary configuration to connect to the LwM2M server. The configuration consists of the credentials and the  LwM2M server URL. Finally, the LwM2M server needs to verify the client's credentials, so it checks this information in the same smart contract as the Bootstrap server.

\subsection{LwM2M clients}
\label{subsec:arch-lwm2m-clients}
LwM2M clients are responsible for making measurements of physical parameters on the environment or infrastructures.  The objective of this project is to provide a system where devices are reliable and where the information generated maintains its integrity and veracity over time, having the minimum repercussion to them.

The register and authentication of the devices are achieved through the Bootstrap mechanism and the Blockchain network.  On the other hand, the integrity of the data of interest is guaranteed by storing them (when needed) on the Blockchain.

\subsection{LwM2M servers and secure API}
\label{subsec:arch-lwm2m-servers}
The objective of LwM2M servers is to allow the monitoring of the environment through the LwM2M clients that are registered on them and their management. As mentioned in subsection \ref{subsec:lwm2m}, the LwM2M protocol allows a quick and easy interaction with clients, making it possible to obtain their information, modify their parameters, initiate an observation and even update the firmware of the device.

LwM2M servers also use Blockchain technology to verify the authenticity of LwM2M clients. Every time a client tries to connect to an LwM2M server, a query is performed to the ClientStore smart contract that stores the security data of all LwM2M clients. This contract is the same one used in the bootstrapping. In this way, a double check is made during client registration, first on the Bootstrap server and later on the LwM2M server itself.

An Application Programming Interface (API) REST is provided on each LwM2M server to allow the access to LwM2M clients and their resources by third-party applications. To make use of this API, it is necessary to have a register with the corresponding permissions. To verify the authenticity of users, the contract UserStore is used, where all registered users are stored.

\subsection{External applications}
\label{subsec:arch-external-apps}
Different applications and users can securely interact with the LwM2M devices deployed in the system. These interactions are carried out through the secure API that LwM2M servers expose. Before that, an authentication process is required where applications and users are verified with their credentials through Blockchain technology.
 
Through the APIs offered by the LwM2M servers, users and applications can list the LwM2M clients connected to each LwM2M server and interact with them through the LwM2M protocol. Moreover, they can also access to the interface offered by the management web UI to register critical information like anomalies when required.

\section{Implementation}
\label{sec:implementation}
\begin{figure*}
    \centering
    \includegraphics [scale=0.8]{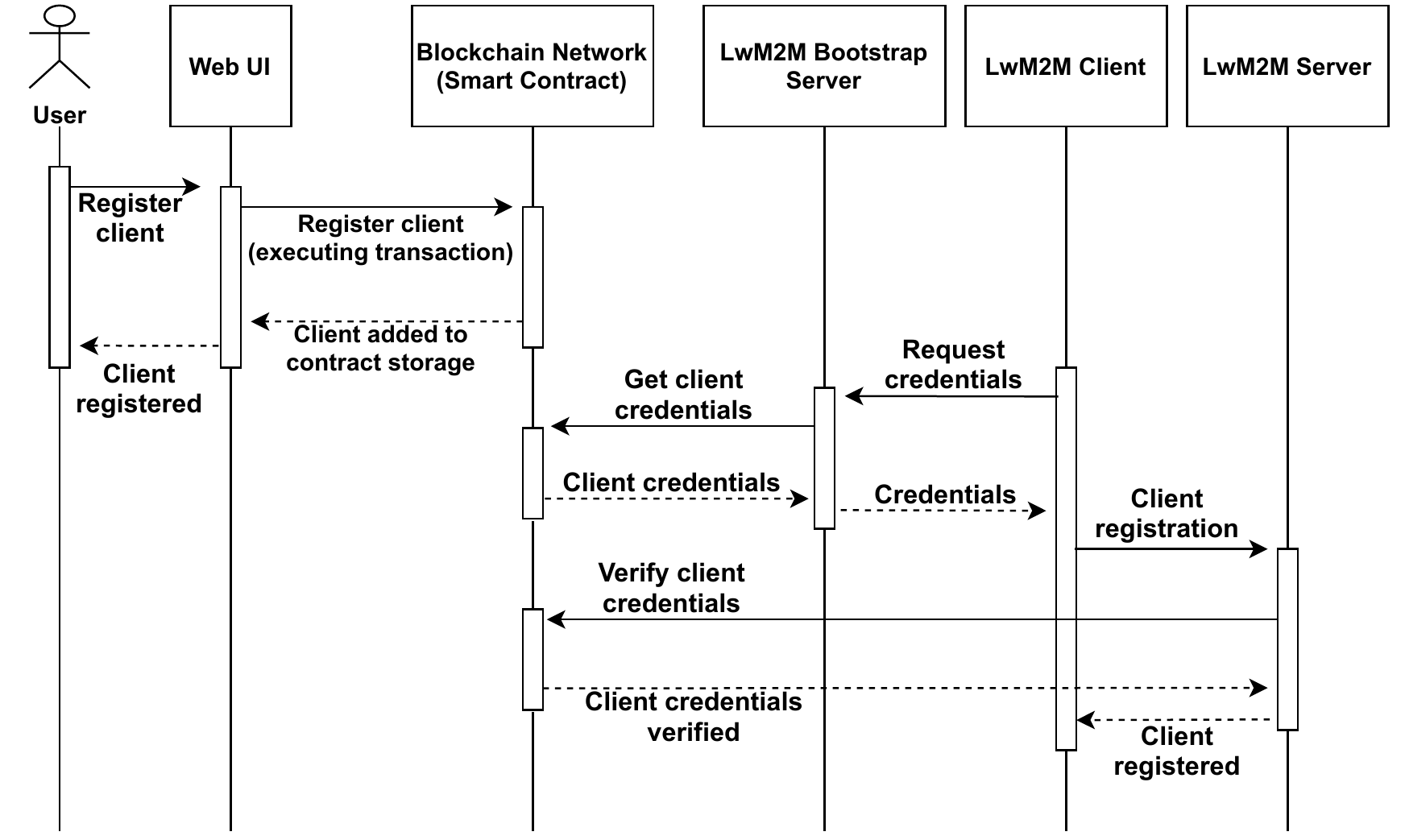}
    \caption{Sequence diagram of an LwM2M client registration in the system and authentication on an LwM2M server}
    \label{fig:seqDiagram}
\end{figure*}

This section describes the implementation of the architecture and components presented in Section \ref{sec:architecture}. This implementation is openly available at our GitHub repository\footnote{https://github.com/ertis-research/lwm2m-blockchain}. To show an overview of the main components and their interactions, a sequence diagram of the LwM2M client registration process is shown in Fig. \ref{fig:seqDiagram}. The diagram begins with a user (admin role) interacting with the web UI, that registers an LwM2M client on the Blockchain network. After this, the LwM2M client requests its credentials to the Bootstrap server and registers itself in an LwM2M server.

Below, the implementation of each component is described.

\subsection{Blockchain network and smart contracts}
\label{subsec:impl-blockchain}
As discussed in subsection \ref{subsec:arch-blockchain}, the Blockchain network should meet two requirements: be public and allow the deployment of smart contracts. Following these constraints, the selected Blockchain network was Ropsten \cite{ropsten}, a testnet of Ethereum which  uses the same algorithm of consensus that Ethereum Mainnet (PoW). In the same way as Ethereum Mainnet, Ropsten needs Ether to deploy smart contracts or to execute transactions, but it is possible to get free Ethers from this network through faucets. A faucet is a mechanism that sends an Ether amount to one account.

To communicate with the Ropsten network, the Infura \cite{infura} project is used. Infura is an infrastructure that allows an easy connection to the Ethereum Mainnet and to all Ethereum testnets including Ropsten through a suite of tools. Once an account and a project are created in Infura, a URL endpoint is provided for each network. These endpoints can be used to deploy smart contracts and to push transactions onto the Ethereum networks.

The proposed system provides three different smart contracts. All of them have been developed using Solidity, an object-oriented, high-level language for implementing smart contracts. The main functionality of each smart contract is described below:
\begin{itemize}
    \item \textit{ClientStore}. All the data about the registered LwM2M clients are saved here. This information includes the client endpoint name, one LwM2M Bootstrap server URL, one LwM2M server URL and client credentials to connect to both servers. The contract has transactions to register new clients, get one or all existing clients and remove a client.
    
    \item \textit{AnomalyStore}. This contract stores critical information such as anomalies, the timestamp when this piece of data was collected and the LwM2M client who provides this information. It is possible to add new critical information entries and get all of them through transactions.
    
    \item \textit{UserStore}. Users management in the Web UI and external applications is carried out by this contract. For each user, a username, an email, a password and a role will be stored. Roles are the same as for the Web UI (see \ref{subsec:arch-management-app}) plus an additional one assigned to the external applications. Available transactions allow to add, get and update users plus validate login attempts.
\end{itemize}

Algorithms \ref{alg:client_store}, \ref{alg:anomaly_store} and \ref{alg:user_store} present the pseudocode of smart contracts \textit{ClientStore}, \textit{AnomalyStore} and \textit{UserStore}, respectively. As shown on these algorithms, Solidity built-in function \textit{revert()} is used when an error occurs (i.e. when trying to add a user that already exists). Besides the main functionality of the smart contracts, each one has an auxiliary function to verify if a client/critical information entry/user exists on smart contract storage. 

\begin{algorithm}
\caption{ClientStore Smart Contract pseudocode}
\label{alg:client_store}
\begin{algorithmic}[1]

\Require{$clients$}\Comment{mapping clients with their configurations}

\Function{addClient}{$client$, $config$} \Comment{Transaction}
\State $client\_exists \gets \textit{clientExists}(client) $
\If{$\neg client\_exists$}
\State $clients[client] \gets config$
\Else
\State $revert()$ \Comment{Built-in Solidity function that reverts a transaction}
\EndIf
\EndFunction

\Function{getClient}{$client$}
\State $client\_exists \gets \textit{clientExists}(client) $
\If{$client\_exists$}
\State $\Return \ clients[client]$
\EndIf
\EndFunction

\Function{getAllClients}{$ $}
\State $\Return \ clients$
\EndFunction

\Function{removeClient}{$client$} \Comment{Transaction}
\State $client\_exists \gets \textit{clientExists}(client) $
\If{$client\_exists$}
\State $\textbf{delete} \ clients[client]$
\Else
\State $revert()$ \Comment{Built-in Solidity function that reverts a transaction}
\EndIf
\EndFunction

\end{algorithmic}
\end{algorithm}

\begin{algorithm}
\caption{AnomalyStore Smart Contract pseudocode}
\label{alg:anomaly_store}
\begin{algorithmic}[1]

\Require{$anomalies$}\Comment{list of anomalies}

\Function{addAnomaly}{$anomaly$} \Comment{Transaction}
\State $n \gets \textit{getNumAnomalies()}$
\State $anomalies[n+1] \gets anomaly$
\EndFunction

\Function{getAllAnomalies}{$ $}
\State $\Return \ anomalies$
\EndFunction

\end{algorithmic}
\end{algorithm}

\begin{algorithm}
\caption{UserStore Smart Contract pseudocode}
\label{alg:user_store}
\begin{algorithmic}[1]

\Require{$users$}\Comment{mapping usernames with users data}

\Function{addUser}{$username, user\_data$} \Comment{Transaction}
\State $user\_exists \gets \textit{userExists}(username) $
\If{$\neg user\_exists$}
\State $users[username] \gets user\_data$
\Else
\State $revert()$ \Comment{Built-in Solidity function that reverts a transaction}
\EndIf
\EndFunction

\Function{getAllUsers}{$ $}
\State $\Return \ users$
\EndFunction

\Function{updateUser}{$username, user\_ data$} \Comment{Transaction}
\State $user\_exists \gets \textit{userExists}(username) $
\If{$user\_exists$}
\State $users[username] \gets user\_data$
\Else
\State $revert()$ \Comment{Built-in Solidity function that reverts a transaction}
\EndIf
\EndFunction

\Function{validateLogin}{$wildcard$} \Comment{Wildcard can be username or email}
\State $user\_exists \gets \textit{userExists}(wildcard) $
\If{$user\_exists$}
\State $\Return \ users[wildcard]$
\EndIf
\EndFunction

\end{algorithmic}
\end{algorithm}

\subsection{Management web UI}
\label{subsec:impl-management-app}
The management web UI follows the traditional client-server model, having an Angular client at the presentation layer (front-end) and a Java/Spring server at the data access layer (back-end). The main difference with the client-server model is the use of a Blockchain instead of a centralized database to store LwM2M client/anomaly/user data.

The communication with smart contracts is carried out through the library web3j\footnote{https://github.com/web3j/web3j}. This library allows the interaction between Ethereum smart contracts and a Java program. With web3j is possible to generate wrapper code to ease the communication with the smart contracts.
Then the server needs to know three parameters to interact with each smart contract:
\begin{itemize}
    \item \textit{URL of a Blockchain network node}: The different queries are made from a Blockchain network node. Infura has been used for this.
    
    \item \textit{Ethereum account}: An Ethereum wallet with ethers is required to execute transactions.
    
    \item \textit{Smart contract address}: After contract deployment, the Blockchain network returns the address where the smart contract is located. If the address number is lost, contract information cannot be retrieved.
\end{itemize}

The previous parameters are configured when the web UI is deployed. The login mechanism implemented for this application makes use of the Blockchain contract \textit{UserStore} to validate credentials through method \textit{validateLogin}. To access the server REST API resources, users must be authenticated. And this authentication is implemented using JSON Web Token (JWT).

JWT is a JSON-based open source standard to create access tokens that allows to secure communication between clients and servers. A JWT process is divided into two phases: authentication and authorization. Firstly, the client sends the user's identity, in this case, username/email and password. Then, the server verifies these credentials using the contract \textit{UserStore} and, if the authentication was successful, a JWT is generated. Depending on the user's role, the token will give access to all resources to admins and restricting LwM2M clients and users management to normal users.

\begin{figure*}[h]
    \centering
    \includegraphics [scale=0.9]{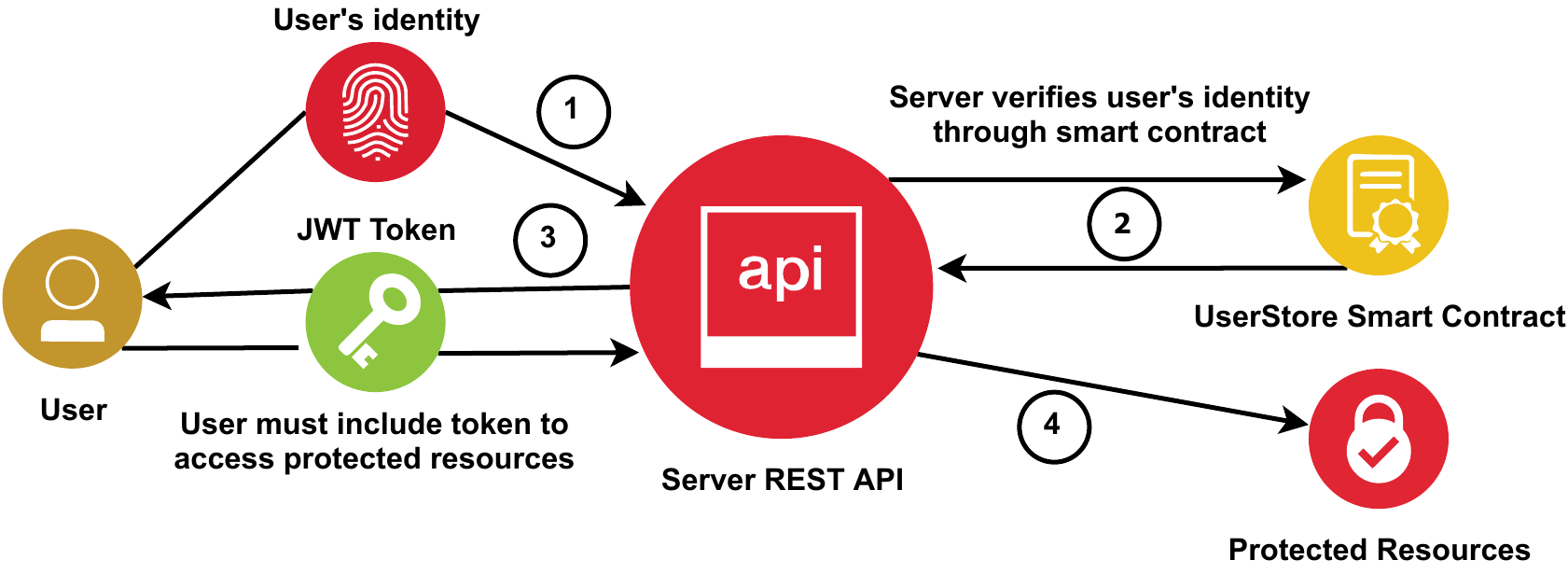}
    \caption{Authentication data flow on management web UI}
    \label{fig:mainAppAuth}
\end{figure*}

For the next requests, users must include this token as a header to access the protected resources. The server decrypts the token and checks if the clients have permissions to access the desired resource. All this data flow is displayed on Fig. \ref{fig:mainAppAuth}.

\begin{figure*}
    \centering
    \includegraphics [scale=0.45]{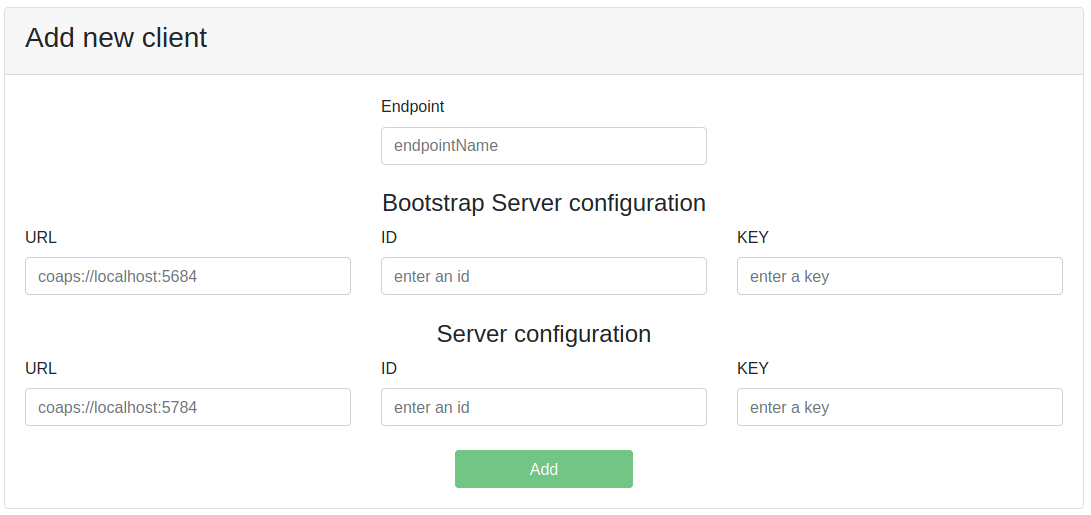}
    \caption{Register an IoT device (LwM2M client) in the Management web UI}
    \label{fig:addclient}
\end{figure*}

Once a user is logged into the management web UI, he/she could perform operations according to his/her role, as for example the secure and reliable register of an LwM2M client into the system (Fig. \ref{fig:addclient}).

\subsection{LwM2M Bootstrap servers}
\label{subsec:impl-lwm2m-bootstrap-servers}

Leshan LwM2M implementation has been used as the base implementation to be extended in this work. Leshan is the most stable and used open-source LwM2M implementation and it provides better documentation and community support than others. Leshan also provides the necessary infrastructure to develop the proposed system in this work, since it implements the four mentioned LwM2M interfaces. Specifically, the way in which Leshan accesses and stores the security information of the different LwM2M clients is modified in such a way that it is consulted in Blockchain and not in a JSON file or in memory as its current implementation does.

To develop the Bootstrap server, a Maven\footnote{https://maven.apache.org/} project was created where the Leshan dependency was included to be able to access all the functionalities and characteristics that Leshan offers. In addition, it was necessary to add the web3j library as a dependency to interact with Ethereum smart contracts through Java as mentioned in subsection \ref{subsec:arch-management-app}.

\subsection{LwM2M clients}
\label{subsec:impl-lwm2m-clients}

LwM2M clients are composed of different objects. Each object has different resources and these objects and resources are identified with an identifier with the form Object/InstanceObject/Resource.  By default, LwM2M clients are created with three mandatory objects: Security, Server and Device. This can be used by any official LwM2M client implementation. This system has been evaluated using the Leshan LwM2M client implementation.

The Security object configures the client to connect to a Bootstrap server using a pre-shared key in order to obtain the necessary credentials to connect to the LwM2M server. The Server object provides the data related to the LwM2M server, but as in this case it connects to a Bootstrap server, it does not have an associated instance.  Lastly, the Device object provides a range of device-related information which can be queried by the LWM2M Server, in addition to a device reboot function.

Along with these three mandatory objects, it is possible to add more objects predefined by OMA such as temperature, humidity sensors, etc., or custom-made objects.

\subsection{LwM2M servers and secure API}
\label{subsec:impl-lwm2m-servers}
A secure REST API, \textit{not available yet in LwM2M Leshan}, has been developed to allow external and authorized applications to access LwM2M servers using the Spring framework. Spring is the most used framework in Java to develop API REST easily. The Spring Boot tool has been used to create the server project since it makes working with dependencies easier.

As mentioned above, clients provide different objects. The LwM2M server needs to load those object models used to know the type of data it expects to receive when it performs a request to a client resource.

\begin{figure*}[h]
    \centering
    \captionsetup{justification=centering}
    \includegraphics [scale=0.4]{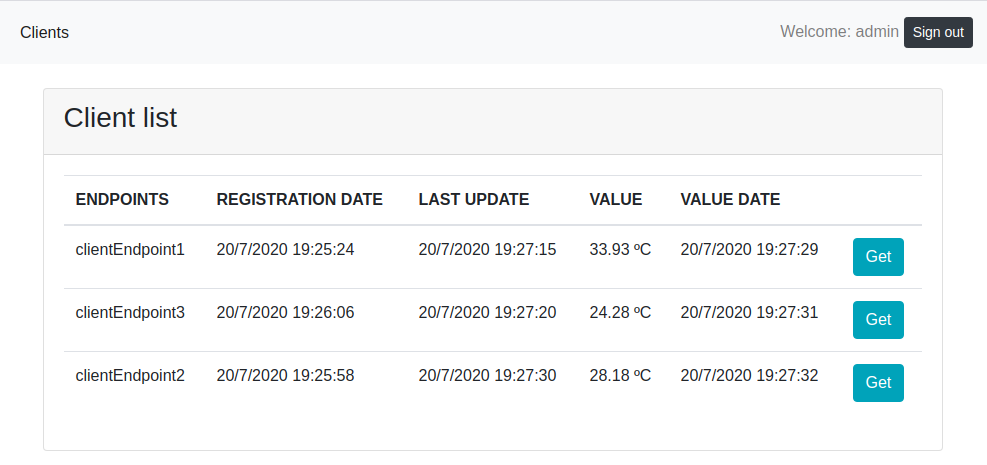}
    \caption{Validation application to interact with LwM2M clients after secure authentication}
    \label{fig:validationapplication}
\end{figure*}

LwM2M servers are configured so that only clients that make secure connections through CoAPS are allowed. LwM2M offers three modes of secure connections: using pre-shared key, with certificate or through Raw Public Key (RPK). Currently, pre-shared key is supported in this project. The verification of the pre-shared key of the clients is carried out in the same way as in the Bootstrap server, consulting the Blockchain network. To do so, the server is deployed and configured with a URL of a Blockchain network node, an Ethereum account with ethers and the address of the deployed smart contract as mentioned in subsection \ref{subsec:impl-management-app}.

In the same way as the management UI, the REST API of each LwM2M server enables authentication using JWT and the deployed smart contract \textit{UserStore}.

\subsection{External applications}
\label{subsec:impl-external-apps}
External and authorized applications can use the REST API offered by LwM2M servers. For this, applications should perform a REST request in any programming language with their credentials and then use the JWT received, the authentication mechanism implemented after authentication, to be able to perform requests and interact with the LwM2M clients through the secure REST API.

For validation purposes, an application has been defined using the popular JavaScript framework Angular. Once a user/application has been authenticated into an LwM2M server, the accessible clients to interact with are displayed (Fig. \ref{fig:validationapplication}). Along with each client, a button has been added to perform a GET request an LwM2M to a temperature resource created in the LwM2M clients.

\section{Evaluation}
\label{sec:evaluation}
This section exhibits the validation carried out on the proposed system and its main functionalities: 1) devices management and authentication; 2) applications and users management and authentication; and 3) critical information management. The average time of 100 tests has been taken for all the cases. The evaluation has been performed in a computer with the following configuration:

\begin{itemize}
    \item \textit{Operating System}: Windows 10 Pro
    \item \textit{CPU}: Intel(R) Core(TM) i7-4790 3.60 GHz
    \item \textit{Storage}: 376 GB SSD
    \item \textit{Memory}: 16 GB RAM
\end{itemize}

The Ropsten test network, deployed publicly around the globe, has also been used to validate the previous smart contracts. When working with Ethereum Mainnet or one of its testnets, it is necessary to set a gas price and a gas limit for transactions. Gas price refers to the cost necessary to perform a transaction on the network while gas limit is the maximum amount of gas that the interested party is willing to spend on a particular transaction. Both parameters are expressed in \textit{Gwei}\footnote{1 \textit{Gwei} = $10
^{-9}$ \textit{Ether}}. These values influence the time taken and the \textit{Gwei} amount spent on each transaction. If both parameters are high, the transactions will be faster and more expensive. On the other hand, if the price of gas is low, the commission earned by miners will not be high, so it will not be an attractive transaction and will take longer to be selected and executed.

To carry out the evaluation, a limit of 4712388 \textit{wei}\footnote{1 \textit{wei} = 10\textsuperscript{-9} \textit{gwei}} and a price per gas of 40 \textit{gwei} has been established. For the selection of these values, the recommended values in \cite{GasPriceEthereum} have been taken into account. It is important to note that these values change frequently, so the times may vary.

\subsection{Devices management and authentication}
\begin{figure}[h]
    \centering
    \captionsetup{justification=centering}
    \includegraphics [scale=0.5]{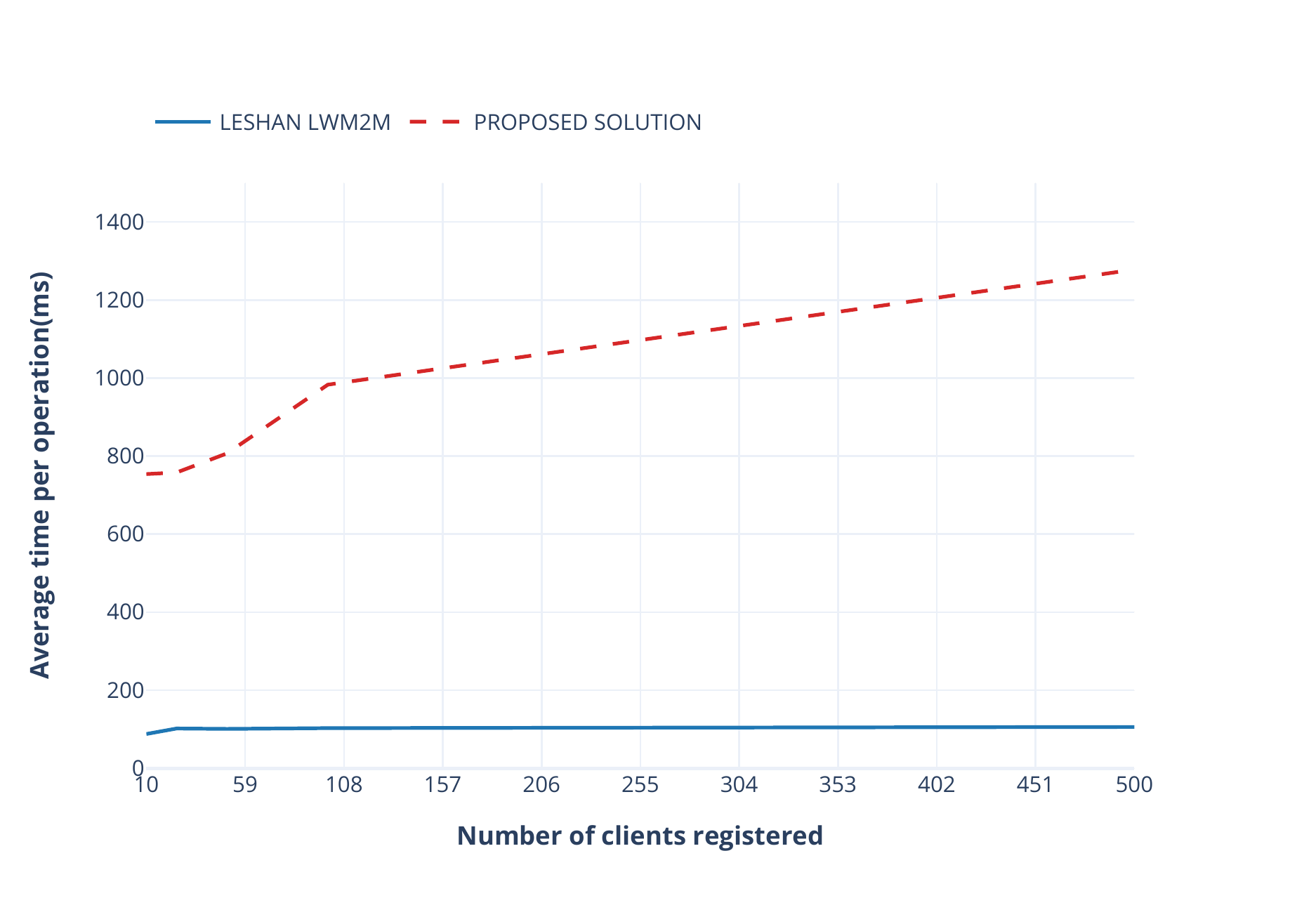}
    \caption{Average elapsed time during client data registration process. In-memory Eclipse Leshan vs proposed Blockchain solution}
    \label{fig:results1}
\end{figure}

The first test performed aims to evaluate the response time when registering  and authenticating new LwM2M clients data in our Blockchain-based system compared with the one offered by the LwM2M implementation of Eclipse Leshan, testing with a different number of clients registered. As Fig. \ref{fig:results1} shows, the total elapsed time on Blockchain is higher. This is due to the overhead provided by Blockchain. However, for this system these times are within an acceptable range since latency is less than 1.3 seconds to register a client information when 500 clients are stored. It is important to note that storing information in memory (as Leshan currently works by default) is meaningless in a real environment. In any case, it should be stored in a database, so response time in Eclipse Leshan should be higher.

\begin{figure}[h]
    \centering
    \captionsetup{justification=centering}
    \includegraphics [scale=0.5]{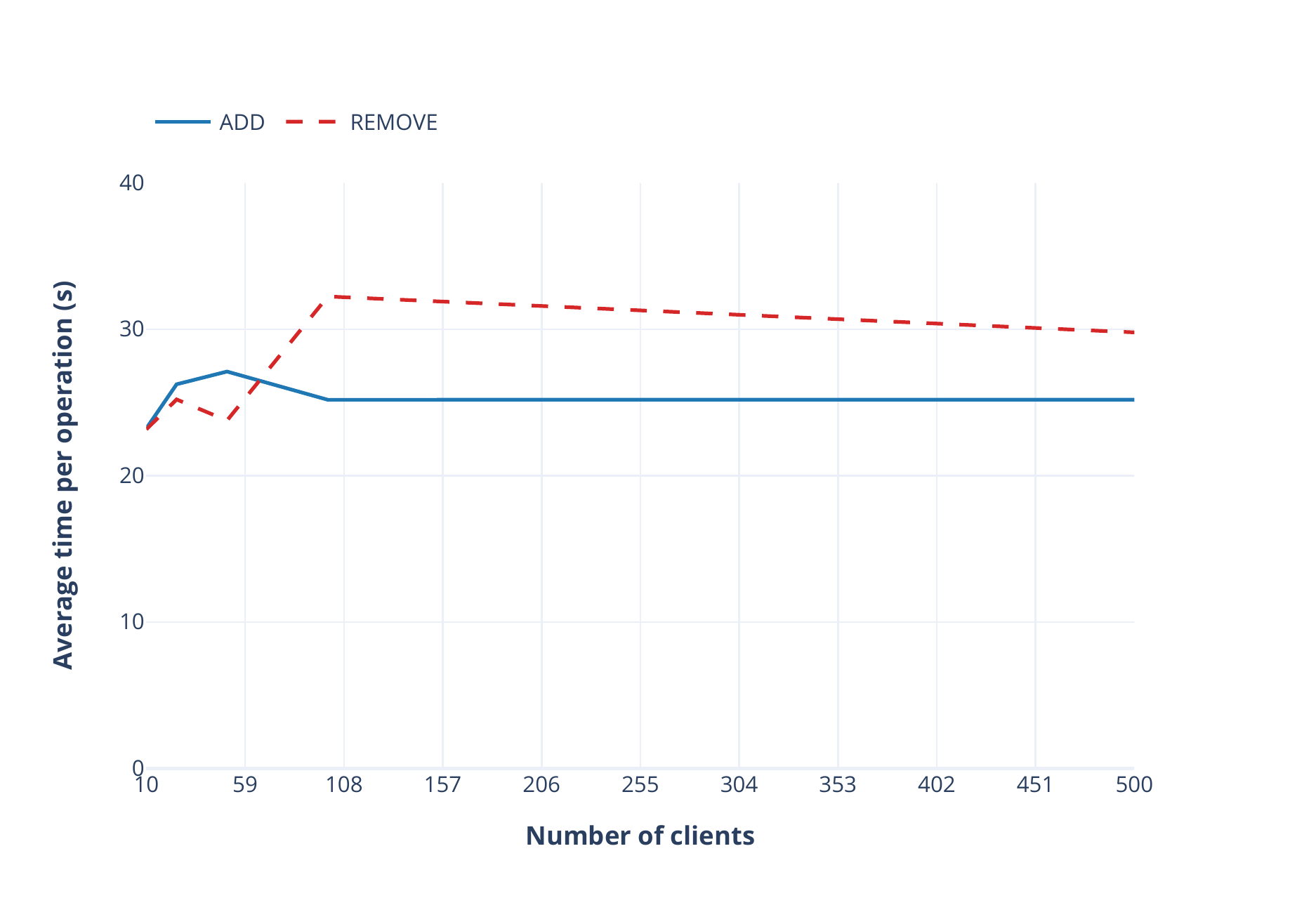}
    \caption{Performance of Add and Remove operations on clients' smart contract}
    \label{fig:results2}
\end{figure}

On the other hand, it has been also evaluated how the number of stored clients affects latency when adding and removing information in the smart contract. Fig. \ref{fig:results2} shows the average times to add and remove clients in the smart contract depending on the total number of clients that are added or removed. In this case, the times obtained have been much higher, around half a minute. This is because this project uses a public Blockchain network, so the execution of transactions (due to the consensus protocol) is slow. In this case, the elapsed time is about tens of seconds. Note that these transactions will not be very frequent since registration of LwM2M clients is usually performed only once for each client.

\subsection{Users/applications management and authentication}

\begin{figure}[h]
    \centering
    \captionsetup{justification=centering}
    \includegraphics [scale=0.6]{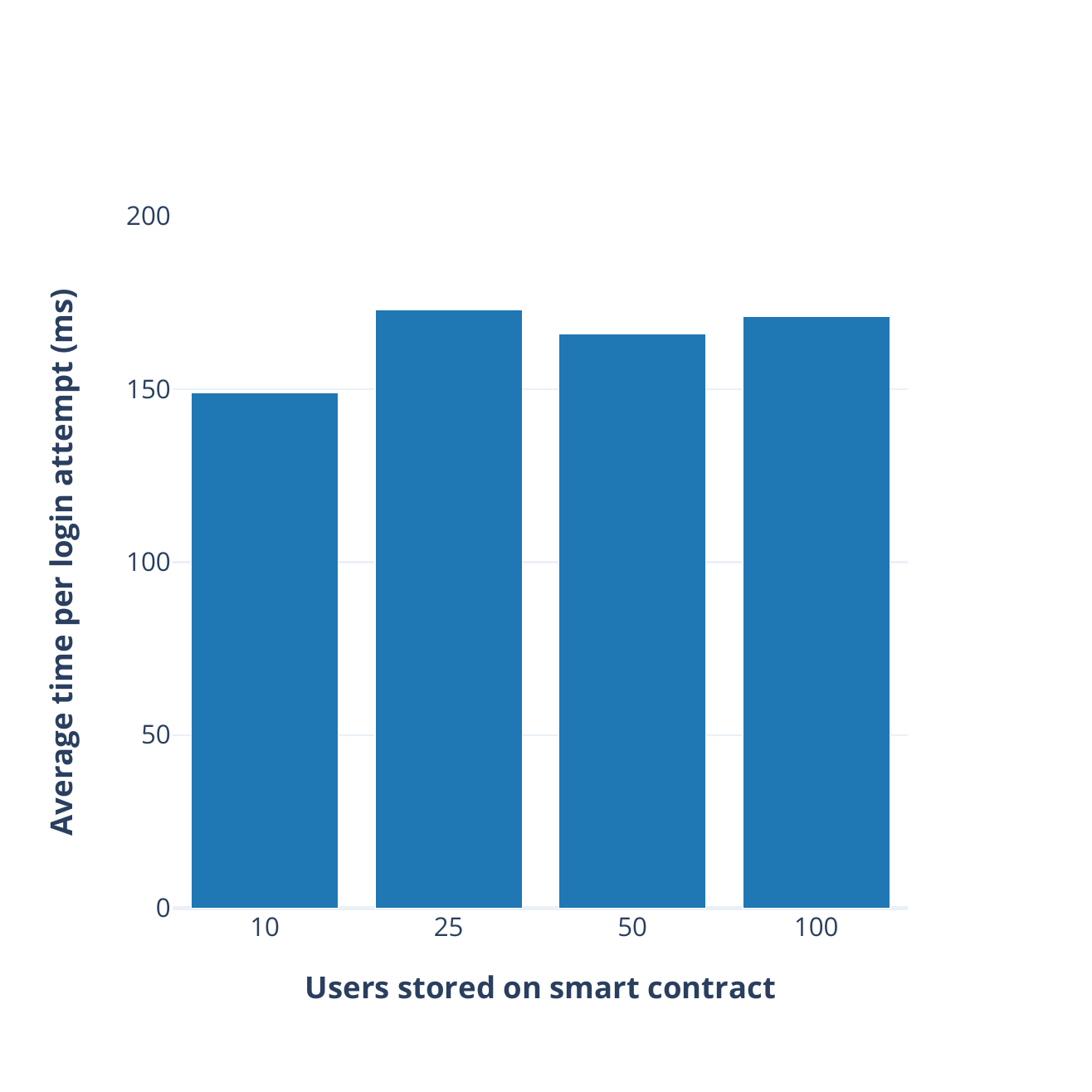}
    \caption{Average elapsed time when authenticating users/applications in the system}
    \label{fig:results6}
\end{figure}

For this smart contract, the latency of the validation process of users'/applications' credentials is evaluated through Blockchain. Fig. \ref{fig:results6} shows the average times obtained depending on the number of users/applications stored in the smart contract. The average query time is around 150-200 ms, a similar response time than the one obtained in the device authentication.

\begin{figure}[h]
    \centering
    \captionsetup{justification=centering}
    \includegraphics [scale=0.5]{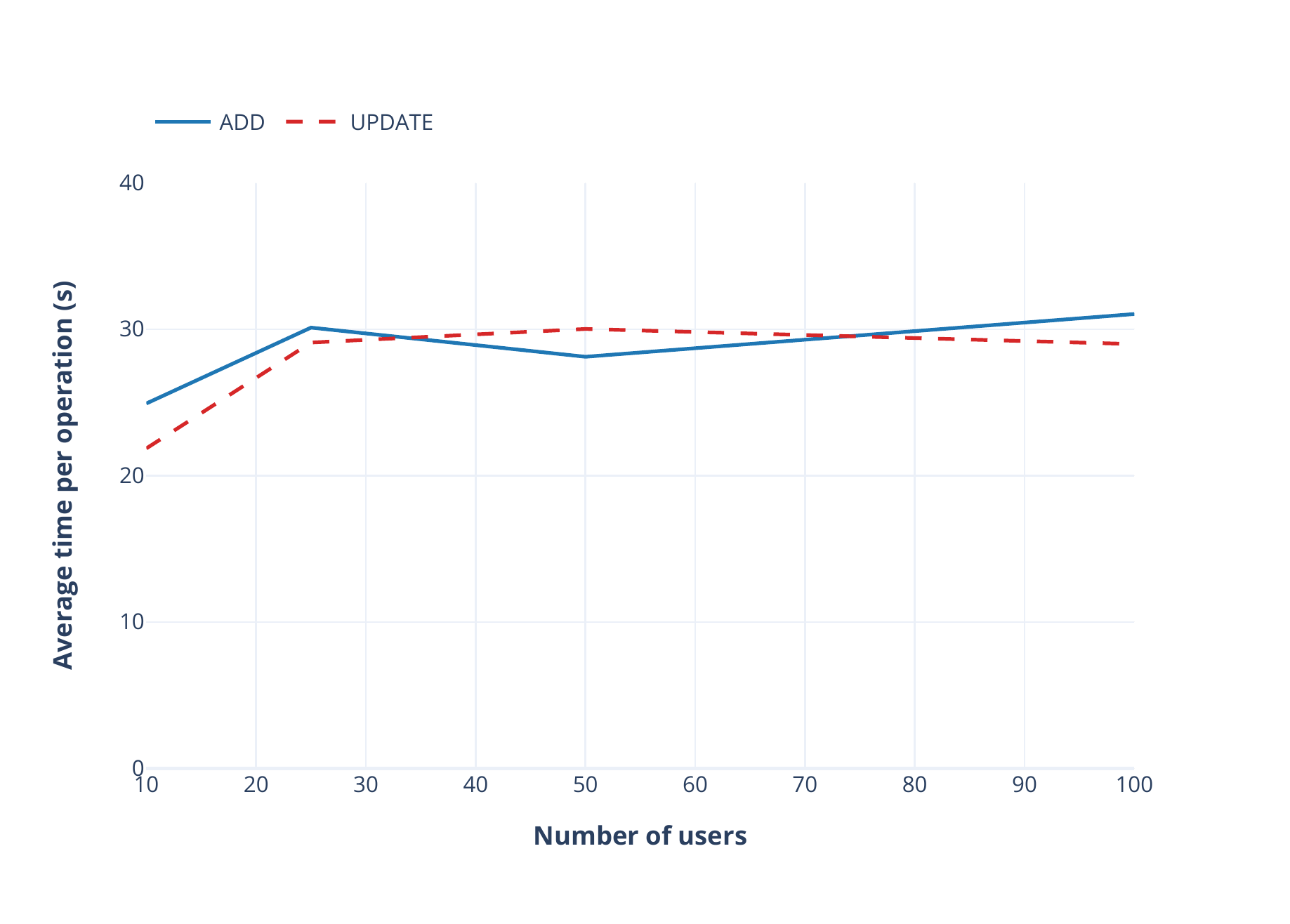}
    \caption{Performance of Add and Update operations on users' smart contract}
    \label{fig:results5}
\end{figure}

On the other hand, the performance is evaluated when adding and modifying applications/users. As in the previous smart contract, when operations that modify the contract are executed, the times obtained are around half a minute as shown in Fig. \ref{fig:results5}. These operations may not be very frequent, so the impact of this latency is not very important in the proposed system.

\subsection{Critical information management}

\begin{figure}[h]
    \centering
    \captionsetup{justification=centering}
    \includegraphics [scale=0.6]{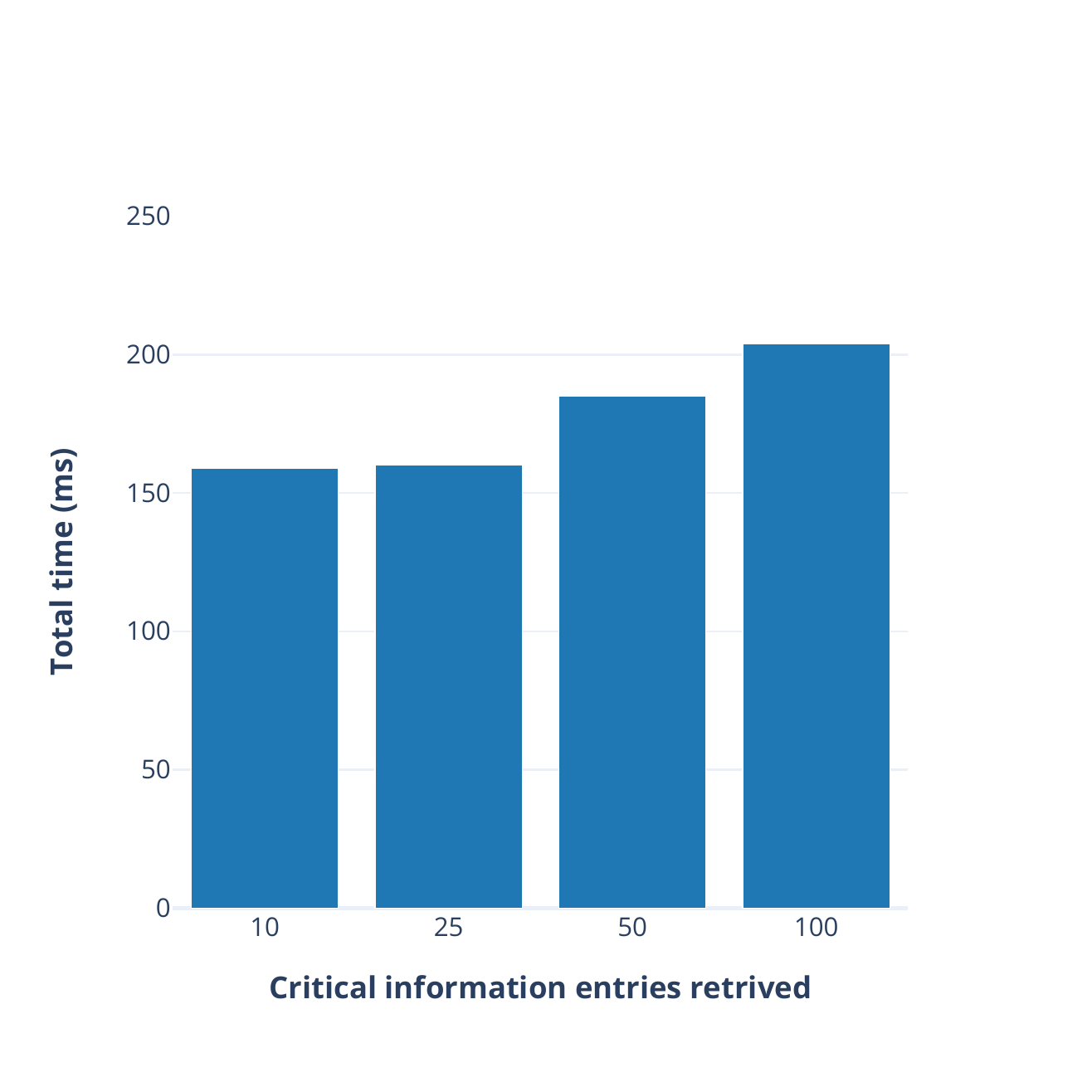}
    \caption{Total time when retrieving all critical information}
    \label{fig:results4}
\end{figure}

Finally, the response time when querying critical information for auditability by authorized users and applications was evaluated. As explained on subsection \ref{subsec:impl-blockchain}, a critical information entry is composed by the timestamp when data were collected, the LwM2M client who provided this information and the critical information itself. Fig. \ref{fig:results4} shows the elapsed time when varying the information stored to evaluate how this also affects latency. As expected, the response time is higher when the information stored is higher due to the features of Blockchain. In any case, the latency is in the order of centiseconds.

\begin{figure}[h]
    \centering
    \captionsetup{justification=centering}
    \includegraphics  [scale=0.6]{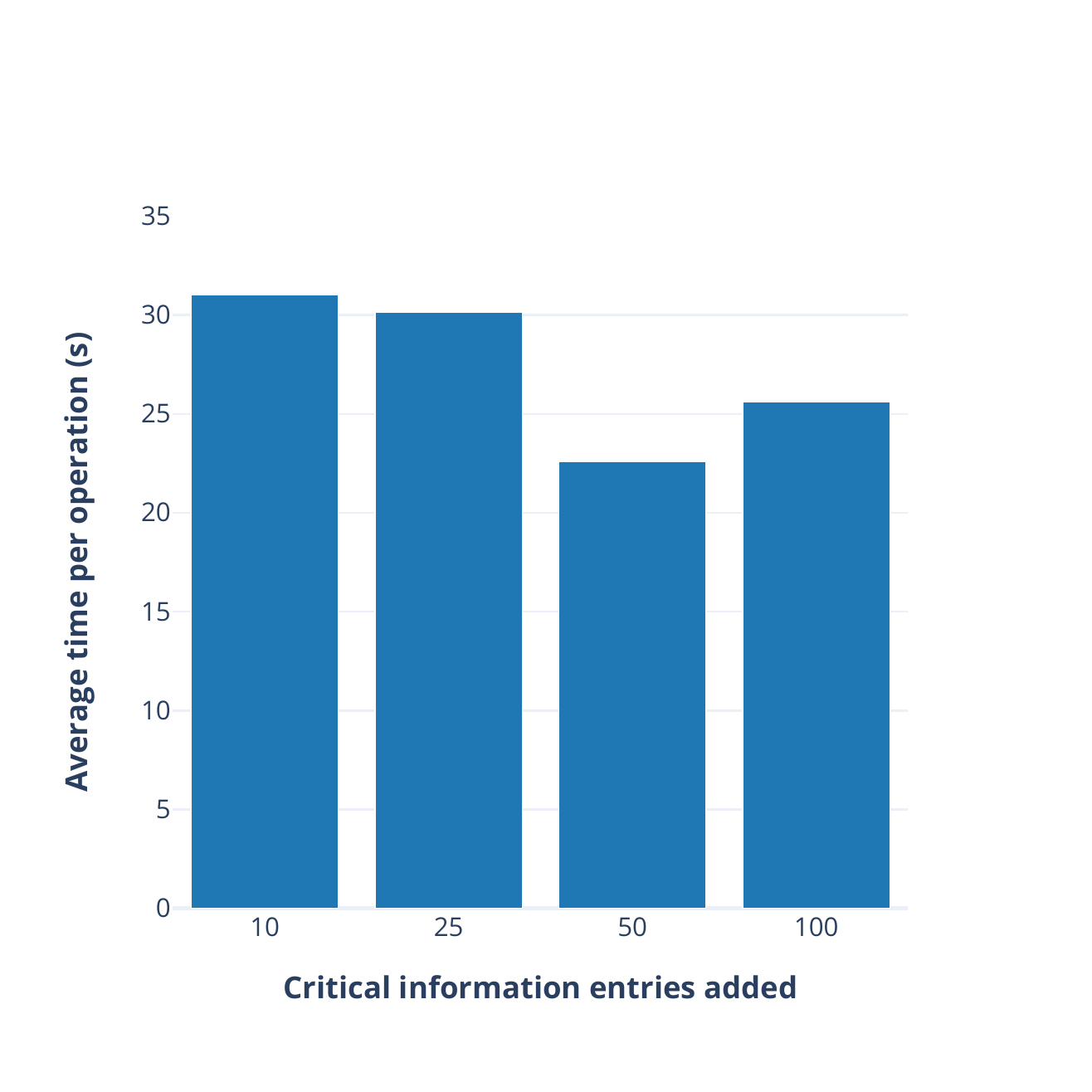}
    \caption{Performance of storing critical information}
    \label{fig:results3}
\end{figure}

Critical information is stored securely and transparently and in an immutable way in the proposed solution for auditability, thereby it should not be updated nor deleted. For this reason, only the response time when adding new information regarding previous information stored was measured (Fig. \ref{fig:results3}). As evaluated in previous contracts, the response time is in the order of tens of seconds when adding new information.

\section{Conclusions and future work}
\label{sec:conclusions-future}
A new integration between the IoT and Blockchain has been proposed in this paper. However, unlike others, this integration focuses on IoT interoperability and provides reliability and auditability to the IoT protocol LwM2M. In particular, the authentication of IoT devices (LwM2M clients) is carried out in the same way as established in the LwM2M standard; however, credentials are securely and transparently stored in Ropsten, a testnet of Ethereum, and are easily registered through a management Web  UI by authorized users. A secure API has been defined to manage the access to external users and applications in LwM2M servers, and thereby to underlying LwM2M clients. Finally, an interface is provided by the management web UI, which can be distributively deployed, to allow applications to register critical information needed for data forensics.

This work mainly focuses on authentication, and even though the access control of applications and users can be managed with this, a more precise access control (e.g., until resource level) is needed in the Industry 4.0. Therefore, new approaches, and a possible integration with the access control list in the LwM2M protocol, will be explored. On the other hand, this work exploits a public Blockchain whose latency can be admissible for device authentication. However, for inter-device communications, this latency may not be acceptable. For this, a hybrid Blockchain considering a private Blockchain will be explored. Finally, a comprehensive evaluation using IoT devices and the application of this work to a real use case are in the roadmap. 

\section*{Acknowledgment}
This work is funded by the Spanish projects RT2018-099777-B-100 (``rFOG: Improving latency and reliability of offloaded computation to the FOG for critical services'') and UMA18FEDERJA-215 (``Advanced Monitoring System Based on Deep Learning Services in Fog'').

\bibliographystyle{IEEEtran}
\bibliography{IEEEabrv,bibliography}

\end{document}